\documentclass[prb,twocolumn,showpacs,tightenlines,10pt]{revtex4}
\usepackage{amssymb}
\usepackage{graphics}
\usepackage{epsfig}

\begin{document}

\title{Anisotropy and large magnetoresistance in narrow gap semiconductor
FeSb$_{2}$}
\author{C. Petrovic, J. W. Kim, S. L. Bud'ko, A. I. Goldman and P. C.
Canfield}
\affiliation{Ames Laboratory and Department of Physics and Astronomy, Iowa State
University, Ames, Iowa 50011}
\author{W. Choe and G. J. Miller}
\affiliation{Ames Laboratory and Department of Chemistry, Ames, Iowa 50011}
\date{\today}

\begin{abstract}
A study of the anisotropy in magnetic, transport and magnetotransport
properties of FeSb$_{2}$ has been made on large single crystals grown from
Sb flux. Magnetic susceptibility of FeSb$_{2}$ shows diamagnetic to
paramagnetic crossover around 100K. Electrical transport along two axes is
semiconducting whereas the third axis exhibits a metal - semiconductor
crossover at temperature T$_{min}$ which is sensitive to current alignment
and ranges between 40 and 80K. \ In H=70kOe semiconducting transport is
restored for T $<$ 300K, resulting in large magnetoresistance [$%
\rho (70$kOe$)-\rho (0)$]$/\rho (0)=2200\%$ in the crossover temperature
range
\end{abstract}

\maketitle

Small gap semiconductors are materials of choice not only as model
electronic systems in materials physics but also in many applications.
Semiconducting compounds often show many phenomena not seen in pure silicon,
such as variety of optical effects, giant magnetoresistances, and ultimately
they can be rather flexible in material design due to possibility for tuning
their fundamental physical properties. Highly anisotropic semiconductors
with directional bands and low dimensional conducting states can provide an
important bridge between bulk and mesoscopic semiconducting materials. One
such material is FeSb$_{2}$.\cite{Nature198Hulliger1081} It represents an
interesting case of a semiconductor where a band of itinerant electron
states originates in the d$_{xy}$ orbitals of the t$_{2g}$ multiplet which
overlap along c-axis of the crystal, distinguishing its loengillate crystal
structure from normal marcasites.\cite{Hulliger}$^{,}$\cite%
{JSSC5Goodenough144} Its magnetic susceptibility is a reminiscent of the one
seen in another narrow gap semiconductor, FeSi, but with very small low
temperature impurity tail in diamagnetic region.\cite{PRL71Schlesinger1748}
In this work we examine the anisotropic magnetic and electronic properties
of FeSb$_{2}$, discuss the possible mechanism for these phenomena and
suggest pathways for further theoretical and experimental work.

Synthesis of large single crystals of FeSb$_{2}$ has allowed us to study the
anisotropy in its magnetic and electrical transport properties. The self
flux method of crystal growth is particularly convenient for the growth of
semiconducting compounds since it does not introduce any additional elements
into the melt which could randomly fill band structure with impurity states.%
\cite{FiskRemeika}$^{,}$\cite{CanfieldFisk}$^{,}$\cite{CanfieldFisher} To
this end, single crystals of FeSb$_{2}$ were grown from an initial
composition of constituents Fe$_{0.08}$Sb$_{0.92}$. The constituent elements
were placed in an alumina crucible and sealed in quartz ampoule. After
initial heating to 1000$%
{{}^\circ}%
C$, the melt was fast cooled \ to 800$%
{{}^\circ}%
C$ in 14h, then slow cooled to 650$%
{{}^\circ}%
C$ where excess Sb flux was removed via decanting. The crystals grew as
silvery rods, their long axis parallel to b crystalline axis.

Room temperature (c.a. 300K) X-ray diffraction data of a single crystal of
FeSb$_{2}$ were collected using a Bruker CCD-1000 diffractometer with Mo K$%
_{\alpha }$ radiation ($\lambda $=0.71073\AA ). The structure solution was
obtained by direct methods and refined by full-matrix least-squares
refinement of F$_{o}^{2}$ using the SHELXTL 5.12 package. Powder X-ray
diffraction spectra are taken with Cu K$_{\alpha }$ radiation in a Scintag
diffractometer. Electrical contacts were made with Epotek H20E silver epoxy.
Resistivity on oriented rectangularly cut single crystals was measured by LR
700 resistance bridge from 1.8 to 300K and in fields up to 70kOe. These
measurements as well as magnetic measurements have been performed in H,T
environment of Quantum Design MPMS-5 and MPMS-7 magnetometers. Magnetic
susceptibility was measured by mounting oriented sample on disk whose
background has been subtracted, in a typical field of 50kOe.

FeSb$_{2}$ crystallizes in marcasite structure similar to rutile (TiO$_{2}$%
), a structure observed primarily for oxides, for example VO$_{2}$. Basic
construction units in both structures are TiO$_{6}$ (FeSb$_{6}$)\ octahedra
that form edge sharing chains along c axis, sharing corners between chains.
The tilt of octahedra in ab plane orthogonal to chain direction
distinguishes the marcasite structure from rutile.

Since phase purity and and questions of exact stoichiometry are important in
semiconductor physics, we have performed a thorough structural analysis. A
crystal with dimensions $0.25\times 0.19\times 0.13$mm$^{3}$ was chosen for
the data set collection. The space groups corresponding to the observed
systematic extinctions are the orthorhombic groups $Pnnm$ and $Pnn2$. We
refined the structure in the $Pnnm$, the centrosymmetric space group of the
two. Lowering symmetry from $Pnnm$ to $Pnn2$ led to no meaningful decrease
in R factor. Crystallographic data taken on single crystal of FeSb$_{2}$ are
in accordance with previously reported, and it is consistent with
orthorhombic marcasite structure with lattice constants a=5.815(4),
b=6.517(5) and c=3.190(2)$\mathring{A}$.\cite{StructureFeSb2} Single crystal
X-ray diffraction measurement showed that site occupancy does not deviate
from ideal FeSb$_{2}$ stoichiometry to within our 1\% resolution limit. In
addition to that, powder X-ray pattern taken on several randomly chosen
samples grown under same conditions was consistent with FeSb$_{2}$ structure
with no additional impurity phases present.

Fig. 1 shows magnetic susceptibility of FeSb$_{2}$ measured along a, b and c
axis of the crystal. It is qualitatively similar to polycrystalline magnetic
susceptibility obtained on crystals grown by a vapor transport technique.%
\cite{JSSC5Fan136} All three directions have similar temperature dependences
but for H $\parallel $ c there is a shift of $\sim $ -1$\times $10$^{-4}$%
emu/mole. The polycrystalline magnetic susceptibility directly measured on
different sample can be estimated by $\chi _{poly}=\frac{1}{3}(\chi
_{a}+\chi _{b}+\chi _{c})$ and is shown in the inset of Fig. 1. It increases
with increase of temperature from low temperature diamagnetic and
temperature independent value of -4$\times $10$^{-5}$ emu/mole (close to
core diamagnetism value of -4.7$\times $10$^{-5}$), passes through a region
of diamagnetic to paramagnetic crossover and becomes paramagnetic at high
temperatures. The crossover temperatures are $\sim $100K for field applied
along a and b axis and $\sim $125K for field applied along c axis.

\begin{figure}
\epsfig{file=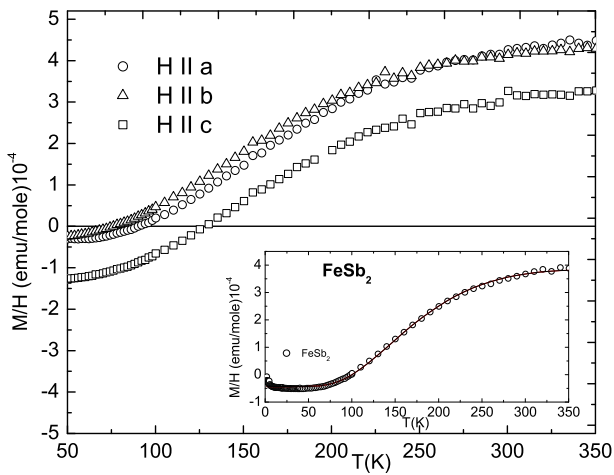,width=0.45\textwidth}
\caption{Magnetic susceptibility of FeSb$_{2}$
single crystal grown by flux method. Inset:\ fit of polycrystalline
susceptibility for low spin to high spin transition. Open circles represent
data taken on different sample. Solid line is fit to a model of low to high
spin transition (see text).}
\end{figure}

Whereas the anisotropy in $\chi (T)$ of FeSb$_{2}$ is relatively
small, the anisotropy in the electrical resistivity $\rho (T)$ is
dramatic (Fig. 2). For the current along either the a and c axis
$\rho (T)$ is semiconducting over the whole temperature range. The
resistivity increases by four orders of magnitude down to lowest
measured temperature of 1.8K (Fig. 2(inset)). From arrhenius plots
of $\rho $(T) curves we can estimate gap values $\Delta _{\rho
}(a,c)\approx 300K$ (Fig. 3 inset), in accordance with previous
results.\cite{JSSC5Fan136}

The b axis transport manifests a metallic behavior above $\sim $40K, with
resistivity ratio (RR)=$\rho $(300K)/$\rho $(40K)=6.3 (extrapolation to T=0
of the high temperature b axis resistivity gives RR=98). Below 40K the b
axis resistivity increases five orders of magnitude, to values comparable to
a and c axis resistivity, and shows activated behavior only below T$_{\min }$%
($\sim $40K\ for optimal current orientation) with activation energy of $%
\Delta _{\rho }(b)\approx 250K$ (Fig. 3 inset). Application of 70kOe along a
and c axis has small influence on resistivity ($\Delta \rho /\rho <0.15$),
but on the other hand crossover temperature region in b-axis resistivity
disappears in this field.

It has been reported that $\rho _{ab}>\rho _{c}$.\cite{JSSC5Fan136} Contrary
to expected, we observe that high conductivity axis is not c, but b axis. It
should be noted though that the observed metallic conductivity in the b axis
electrical transport as well as the T$_{\min }$ are very sensitive to
current misalignment. The effect of deliberate small misalignment in current
path along b axis in ab plane is shown in Fig. 3. RR above T$_{\min }$ for
sample 1 can be changed by a factor of two and T$_{\min }$ itself can be
shifted 30K up in temperature.

\begin{figure}
\epsfig{file=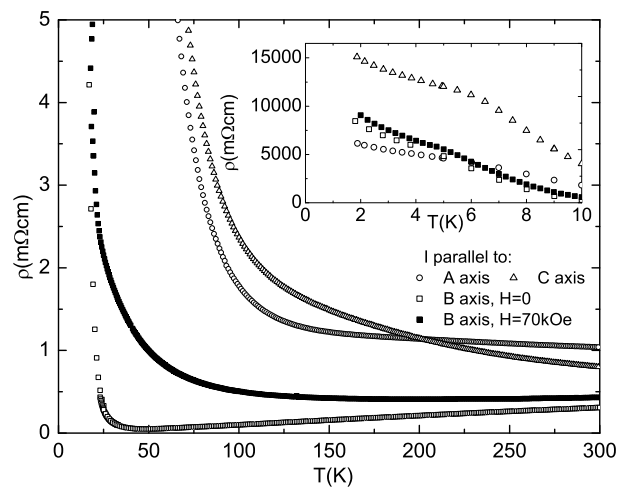,width=0.45\textwidth}
\caption{Anisotropy in electrical transport of FeSb$_{2}$. Inset shows low
temperature resistivity with clear contribution of impurity states below 5K.}
\end{figure}

\begin{figure}
\epsfig{file=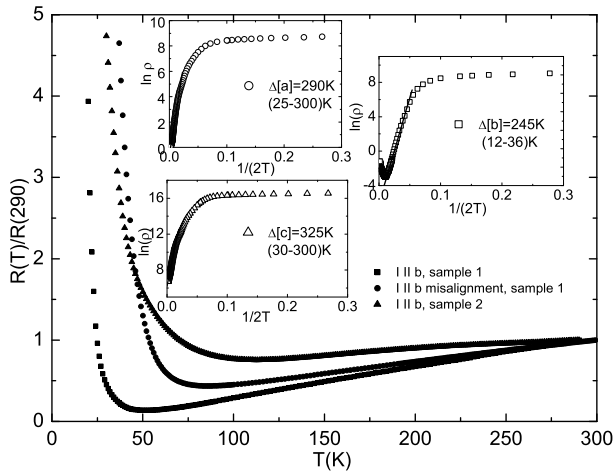,width=0.45\textwidth}
\caption{Crossover temperature region of
b axis electrical transport. Note substantial influence of deliberate current
misalignment ($\sim $10\%) in ab plane for sample 1. Inset shows activated
behavior of resistivity for all three crystalographic directions.}
\end{figure}

As shown in Fig. 2, an applied
field enhances b-axis resistivity near T$_{\min }$ leading to a large
magnetoresistance. The 70kOe magnetoresistance is temperature dependant, and
it has a sharp maximum $\Delta \rho /\rho =22$ in the crossover region (Fig.
4).

\begin{figure}
\epsfig{file=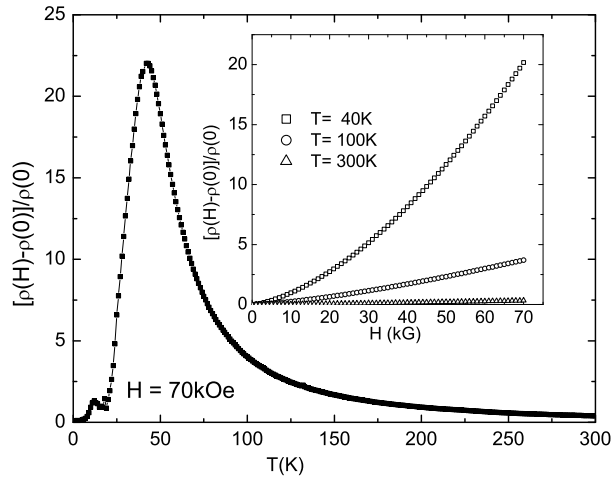,width=0.45\textwidth}
\caption{Temperature dependant b
axis magnetoresistance. Inset shows magnetic isotherms around T$_{\min }$
(40K) and in the paramagnetic region (200K and 300K).}
\end{figure}

Magnetic isotherms (Fig. 4 inset) show H$^{\alpha \text{ }}$%
dependence where $\alpha =1.5-1.7,$ a value smaller than $\alpha =2$
expected for a simple one-carrier system with energy independent carrier
relaxation time ($\Delta \rho (H)/\rho _{0}=\mu ^{2}H^{2}$ where $\mu $ is
is the carrier mobility).

The marcasite-type FeSb$_{2}$ has been classified as a semimetal or narrow
gap semiconductor\cite{Nature198Hulliger1081}$^{,}$\cite{JSSC5Fan136} in
which both valence and conduction band are derived from d-like states.\cite%
{JSSC5Goodenough144}$^{,}$\cite{SSC33Goncalves63} We rationalize our
observation of anisotropy in its physical properties within the framework of
temperature induced transitions within the 3d multiplet.

In the orthorhombic marcasite structure Fe (cation) is surrounded by a
deformed Sb (anion) octahedra. These octahedra then share edges along
c-axis. The edge sharing octahedra form chains parallel to c axis causing
overlap of d$_{xy}$ atomic orbitals. As opposed to normal marcasites with
filled d$_{xy}$ orbitals and a c/a ratio between 0.73-0.75, loellingites
with d$^{2}$ and d$^{4}$ cations have c/a ratio between 0.50-0.53 and empty d%
$_{xy}$ orbital\cite{Nature198Hulliger1081}$^{,}$\cite{Hulliger}. Based on
the scheme given by Goodenough, t$_{2g}$ orbitals are further split in two
lower lying ($\Lambda $) orbitals associated with d$_{xz}$ and d$_{yz}$, and
a higher lying ($\Xi $) orbital associated with d$_{xy}$ which create $\Xi $
band of itinerant electron states due to their overlap\cite%
{JSSC5Goodenough144}$^{,}$\cite{labeling}. Starting at low temperatures, FeSb%
$_{2}$ is a diamagnetic semiconductor, as expected for S=0 low spin d$^{4}$
ground state (t$_{2g}^{4}$, S=0) where low energy $\Lambda $ orbitals are
filled with electrons with opposite spin due to crystalline fields which are
larger than Hund's rule spin pairing energy. We performed analysis of the
thermal excitation from ground state nonmagnetic (S=0)\ to paramagnetic
excited (S$\neq 0$) state: it results in a change of magnetic susceptibility
$\chi (T)=Ng^{2}\mu _{B}^{2}\frac{J(J+1)}{3k_{B}T}\frac{2J+1}{2J+1+\exp
(\Delta _{\chi }/k_{B}T)}$ where J=S and $\Delta _{\chi }$ is susceptibility
gap.\cite{PR160Jaccarino476} A fit to polycrystalline average of our data
over the whole temperature range for fixed g=2 (Fig. 1 inset) describes well
behavior of FeSb$_{2}$ and yields $\Delta _{\chi }\approx 546K$, S=0.59 for
spin value in Curie constant and $\chi _{0}$=-4$\times 10^{-5}$ (emu/mole).

One possible explanation for enhanced conductivity in paramagnetic state and
its anisotropy is the population of the $\Xi $ band of itinerant states
induced by thermal depairing on low energy $\Lambda $ orbitals. As the
temperature is raised, some of 3d electrons are thermally excited to $\Xi $
band responsible for conduction, while electrons in localized $\Lambda $ or e%
$_{g}$ orbitals are responsible for temperature induced paramagnetic moment,
as seen in magnetic susceptibility which shows significant enhancement above
100K. Delocalization in this scenario is connected with transition within $%
t_{2g}$ multiplet and it is possible that it occurs at lower temperatures
than the transition to higher lying e$_{g}$ orbital which explains T$_{\min
} $ as low as 40K for current applied along b axis in diamagnetic state. We
also note the difference between susceptibility ($\Delta _{\chi }$) and
resistivity ($\Delta _{\rho }$) gaps, indicating that the gap relevant for
conductivity is smaller than gap relevant for the susceptibility, an
observation which is not in contradiction with above description. A possible
difference between gaps in charge and spin excitation channels has been also
observed in some samples of FeSi.\cite{PRB56Paschen12916}

The magnetic susceptibility of FeSb$_{2}$ is reminiscent of $\chi (T)$ data
seen in FeSi, albeit with diamagnetic susceptibility at low temperature and
a much smaller tail below 5K. Apart from the "free-ion"-like model of
localized electrons described above, the model of metallic paramagnetism by
Jaccarino \textit{et al.} has been invoked to apply magnetic susceptibility of FeSi.%
\cite{PRL71Schlesinger1748}$^{,}$\cite{PR160Jaccarino476}$^{,}$\cite%
{PR18Wertheim89}$^{,}$\cite{PRB51Mandrus4763} Attempts to interpret the
magnetic susceptibility of FeSb$_{2}$ within this model of two narrow bands
with rectangular and constant density of states of width W separated by
energy gap E$_{g}$ did not produce meaningful fitting parameters. More
refined analysis with a different band shape and photoemission spectroscopy
measurements could offer a more definite statement about validity of narrow
band Kondo insulator - like description of this material. Moreover, since
the difference in $\Delta _{\chi }$and $\Delta _{\rho }$ seen in FeSi was
explained in the framework of metallic paramagnetism by invoking the
existence of indirect (smaller) energy gap responsible for transport and
direct (larger) gap of the same width for both spin and charge excitations,
possible Kondo insulator - like features in FeSb$_{2}$ deserve further study.%
\cite{PRB56Paschen12916}

The large magnetoresistance (MR) seen in FeSb$_{2}$ for I\ $\parallel $ b
axis ($\sim $2250\% at T$_{\min }$ and $\sim $32\% at T=300K in H=70kOe) is
comparable in magnitude to MR seen in giant magnetoresistance (GMR)
materials such as manganate perovskites.\cite{Science264Jin413}$^{,}$\cite%
{RMP73Salomon583} The spin disorder scattering mechanism of MR does not seem
to be a viable mechanism in this material. One possible, but speculative,
explanation of the large magnetoresistance phenomenon can be found in
analogy with the extraordinary magnetoresistance (EMR) seen in non-magnetic
semiconductors with embedded metallic inhomogenieties.\cite{APL72Thio3497}$%
^{,}$\cite{PRB57Thio12239} Since the $\Xi $ band of conducting states is
highly directional in real space, (our measurement in Fig. 3 also is
consistent with this interpretation), it can act as a region of metallic
conductivity in a semiconducting environment, short-circuiting the most of
applied current passing through it. In the simplest picture of isotropic
conductivity, the single band carrier mobility is $\mu =R_{H}\sigma $. By
including scattering time $\tau $ through general relation R$_{H}$=-$\omega
_{c}\tau /\sigma B$, we obtain $\mu B\sim \omega _{c}\tau $. Large positive
magnetoresistance is then a consequence of the large mobility of the
carriers in the itinerant $\Xi $ band since even modest fields could enhance
value of $\omega _{c}\tau $. The steep rise of the Hall constant \ below
120K seen in Ref. 7 holds promise of reaching R$_{H}\sim $10$^{-1}$cm$^{3}$%
/C around T$_{\min }$=40K for b axis resistivity. Taking $\rho (T_{\min
})\sim 50\mu \Omega $cm from our measurement, we estimate $\mu (T_{\min
})\sim 2000cm^{2}/Vs$, comparable to high mobility values found in antimony
based materials with skutterudite structure.\cite{PRB51Morelli9622}$^{,}$%
\cite{NASATech} Hence, the condition H $>$ 1/$\mu $ is satisfied
for fields of the order of 50kOe. Strong magnetoresistance
therefore is likely to have its origin in band effects, and the
above description is further supported with Kohler's rule $\Delta
\rho /\rho _{0}$ and H/$\rho _{0}$ curves which fall on the single
manifold (not shown) in the metallic region of b axis conductivity
from 40K to 300K. Measurement of the Hall coefficient at low
temperature would be useful to clarify this issue as well as
further crystallographic studies and band structure calculations
for elucidating the orientation of $\Xi $ band. In addition,
neutron scattering experiments could offer decisive information
about thermally induced paramagnetism. Further study may explain
physics contained in FeSb$_{2}$ in single-electron picture, but on
the other hand it might turn into a playground for many body
effects in 3d material with anisotropic crystal and possible
electronic structure. Since narrow gap semiconductors are
important ingredients in optoelectronic devices for both civilian
and military use, further study and tuning of FeSb$_{2}$
properties deserves some attention.

We thank Zachary Fisk, Vladimir Kogan and Maxim Dzero for useful
discussions. This work was carried out at Ames Laboratory, which is operated
for the U.S. Department of Energy by Iowa State University under Contract
No. W-7405-82. This work was supported by the Director for Energy Research,
Office of Basic Energy Sciences of the U.S. Department of Energy.


\begin{thebibliography}{99}
\bibitem{labeling} In labeling scheme given by Goodenough $\Lambda $ (Ref.
8) orbitals are labeled "b", and higher lying $\Xi $ orbital is labeled "a$%
_{\parallel }$". We use different notation in order to avoid confusion with
crystalline axes orientation.
\end{thebibliography}
\end{document}